\documentclass[prd,twocolumn,aps,showpacs,nofootinbib,nobibnotes,superscriptaddress,preprintnumbers]{revtex4-1}
\usepackage{epsfig}
\usepackage{graphicx}
\usepackage{bm}
\usepackage[usenames, dvipsnames]{color}
\usepackage{dcolumn}   
\usepackage[spanish,english]{babel}
\usepackage{bm}     
\usepackage{bbm}       
\usepackage{amssymb}  
\usepackage{amsmath}
\usepackage{latexsym}
\usepackage{float}
\usepackage{ifthen}
\usepackage{caption,subfig}
\usepackage{enumerate}
\usepackage{url}
\usepackage{caption,subfig}
\usepackage{amsopn}
\usepackage{hyperref}

\usepackage{amsfonts}
\usepackage{multirow}
\usepackage{array}
\usepackage{booktabs}
\usepackage{rotating}

\usepackage{ulem}
\def\clap#1{\hbox to 0pt{\hss#1\hss}}

\def\bea{\begin{eqnarray}}
\def\eea{\end{eqnarray}}
\def\be{\begin{equation}}
\def\ee{\end{equation}}
\def\mpl{M_{\rm P}}

\begin{document}

\title{Dark Energy in the Swampland}

\author{Lavinia Heisenberg} \email{lavinia.heisenberg@eth-its.ethz.ch}
\affiliation{Institute for Theoretical Studies, ETH Zurich, 
 Clausiusstrasse 47, 8092 Zurich, Switzerland}
 
\author{Matthias Bartelmann} \email{bartelmann@uni-heidelberg.de}
\affiliation{Universit\"at Heidelberg, Zentrum f\"ur Astronomie, Institut f\"ur Theoretische Astrophysik, Germany}

\author{Robert Brandenberger} \email{rhb@hep.physics.mcgill.ca}
\affiliation{Physics Department, McGill University, Montreal, QC, H3A 2
T8, Canada}

\author{Alexandre Refregier} \email{alexandre.refregier@phys.ethz.ch}
\affiliation{Institute for Particle Physics and Astrophysics, Department of Physics, ETH Zurich, Wolfgang-Pauli-Strasse 27, 8093, Zurich, Switzerland}

\date{\today}

\begin{abstract}
In this Letter, we study the implications of string Swampland criteria for dark energy in view of ongoing and future cosmological observations. If string theory should be the ultimate quantum gravity theory, there is evidence that exact de Sitter solutions with a positive cosmological constant cannot describe the fate of the late-time universe.  Even though cosmological models with dark energy given by a scalar field $\pi$ evolving in time are not in direct tension with string theory, they have to satisfy the Swampland criteria $|\Delta\pi|<d\sim\mathcal{O}(1)$ and $|V'|/V>c\sim\mathcal{O}(1)$, where $V$ is the scalar field potential. In view of the restrictive implications that the Swampland criteria have on dark energy, we investigate the accuracy needed for future observations to tightly constrain standard dark-energy models. We find that current 3-$\sigma$ constraints with $c \lesssim 1.35$ are still well in agreement with the string Swampland criteria. However, Stage-4 surveys such as Euclid, LSST and DESI, tightly constraining the equation of state $w(z)$, will start putting surviving quintessence models into tensions with the string Swampland criteria by demanding $c<0.4$. 
We further investigate whether any idealised futuristic survey will ever be able to give a decisive answer to the question whether the cosmological constant would be preferred over a time-evolving dark-energy model within the Swampland criteria. Hypothetical surveys with a reduction in the uncertainties by a factor of $\sim20$ compared to Euclid would be necessary to reveal strong tension between quintessence models obeying the string Swampland criteria and observations by pushing the allowed values down to $c<0.1$.  
In view of such perspectives, there will be fundamental observational limitations with future surveys.

\end{abstract}


\maketitle


\section{Introduction}

Einstein's theory of General Relativity (GR) is still the standard effective field theory for the gravitational interaction below the Planck scale. Having survived a multitude of empirical tests in a wide range of scales, it continues to stand out as the most compelling candidate theory. It thus constitutes the bedrock upon which all of our effective field theories of gravity are constructed (see e.g. \cite{Heisenberg:2018vsk} for a recent review). However, some tenacious challenges remain, concerning in particular its UV completion into a quantum gravity theory and the necessity of enigmatic ingredients in form of dark energy and dark matter. A prevailing view is that GR could be just the low energy limit of the more fundamental string theory, and that some of the IR/UV problems could be cured by particularities of the UV completion.

String theory refers to physical models that, instead of describing elementary particles as point-like objects in space-time in the familiar models of quantum field theory, introduces strings as fundamental objects. Originally, string theories were used to describe the strong interaction, where the gluons were interpreted as spatially extended strings between the quarks. String theory has received much interest, however this time as a candidate for a unified theory combining the standard model of particle physics with gravity. Should string theory be the ultimate theory of quantum gravity, the question immediately arises whether the effective field theories of gravity known to us can naturally be embedded into string theory. In this context, the Swampland \cite{Vafa2} emerges as the inhabitable landscape of field theories that are incompatible with quantum gravity.

The implications of these Swampland criteria are tremendous (see e.g. \cite{recent} for some discussions). There is evidence that stable de-Sitter vacua in critical string theory do not exist (see e.g. \cite{Obied:2018sgi,noLambda} for some recent references) \footnote{On the other hand, specific constructions of at least metastable de Sitter vacua arising from string theory have been proposed using effective field theory techniques.}.
If the late-time universe is dominated by a dark energy scalar field evolving with time, string-theory criteria can be used to constrain these dark-energy models. 
In this Letter, we discuss the implications of the string Swampland criteria for scalar field dark energy. 

\section{String Swampland}

We will consider General Relativity in the presence of a quintessence field $\pi$ as our effective field theory with the action
\begin{equation}
\mathcal{S}=\int d^4x \sqrt{-g}\left\{ \frac{\mpl^2}{2}R-\frac12\partial_\mu\pi\partial^\mu\pi-V(\pi) \right\}
\end{equation}
to be added to the action $\mathcal{S}_{\rm m}$ of the standard matter fields. Introducing the dynamical variables $x=\dot\pi/(\sqrt{6}H\mpl)$ and $y=\sqrt{V}/(\sqrt{3}H\mpl)$, we can bring the background equations of motion of the scalar field and the metric into the autonomous form \cite{Tsujikawa:2013fta}
\begin{align}\label{EOMquintessence_a}
\frac{dx}{dN}&=-3x+\frac{\sqrt{6}}{2}\lambda y^2+\frac32x\mathcal{F}\;,\\
\frac{dy}{dN}&=-\frac{\sqrt{6}}{2}\lambda xy+\frac32y\mathcal{F}\;,
\label{EOMquintessence_b}\end{align}
where we have defined $\lambda=-\mpl V'/V$, $N=\ln a$ and $\mathcal{F}=\left[(1-w_m)x^2+(1+w_m)(1-y^2)\right]$, while $w_m$ represents the equation-of-state parameter of the standard matter fields. The equation-of-state parameter of the scalar field is $w=(x^2-y^2)/(x^2+y^2)$. The relative potential derivative $\lambda$ satisfies
\begin{equation}
\frac{d\lambda}{dN}=-\sqrt{3(1+w)(x^2+y^2)}\left(\frac{VV''}{V'^2}-1\right)\lambda^2\,.
\end{equation}

This effective field theory admits solutions modelling an accelerated universe relevant for dark energy. However, these solutions have to satisfy certain criteria in order not to end up in the Swampland. The two Swampland criteria on an effective field theory consistent with string theory are that, given a point in field space,
\begin{itemize}  
  \item the range traversed by a scalar field is bounded by $|\Delta\pi|<d\sim\mathcal{O}(1)$ in reduced Planck units \cite{Ooguri:2006in}; and
  \item the derivative of the scalar-field potential has to satisfy the lower bound $|V'|/V>c\sim\mathcal{O}(1)$ \cite{Obied:2018sgi}.
\end{itemize}
If the field traverses a larger distance, then one leaves the domain of validity of the effective field theory (new string states become massless). Within a domain over which the field evolves, the second condition then needs to be satisfied.
The second of these two criteria will be primarily relevant and interesting in view of dark-energy applications. As it was shown in \cite{Agrawal:2018own}, the constant $\lambda$ case with $\lambda=c$ is the least constrained trajectory and therefore we will focus on this case here. Since we are interested in the dark energy dominated epoch we will assume $w_m\sim0$. By way of the relations $w_{\rm eff}+1=-2\dot{H}/(3H^2)$ and $w=(\dot\pi^2-2V)/(\dot\pi^2+2V)$, the condition $|V'|/V>c$ translates into $w+1>0.15c^2$ in reduced Planck units \cite{Agrawal:2018own}. In this Letter, we will extend the work of Agrawal et al.\  \cite{Agrawal:2018own} and investigate the Swampland criteria in view of prospectively achievable observational uncertainties.

\section{Current bounds}

Following \cite{Agrawal:2018own}, we apply the observational bounds from SNeIa, CMB, BAO and $H_0$ measurements derived in \cite{Scolnic:2017caz}. We use the 1- and 2-$\sigma$ contours of Fig.\ 21 of \cite{Scolnic:2017caz} on the Chevallier-Polarski-Linder (CPL) parametrisation \cite{Chevallier:2000qy} and translate them to an upper bound on the reconstructed equation of state $w(z)$ as a function of redshift.

\begin{figure}[h!]
  \includegraphics[width=\hsize]{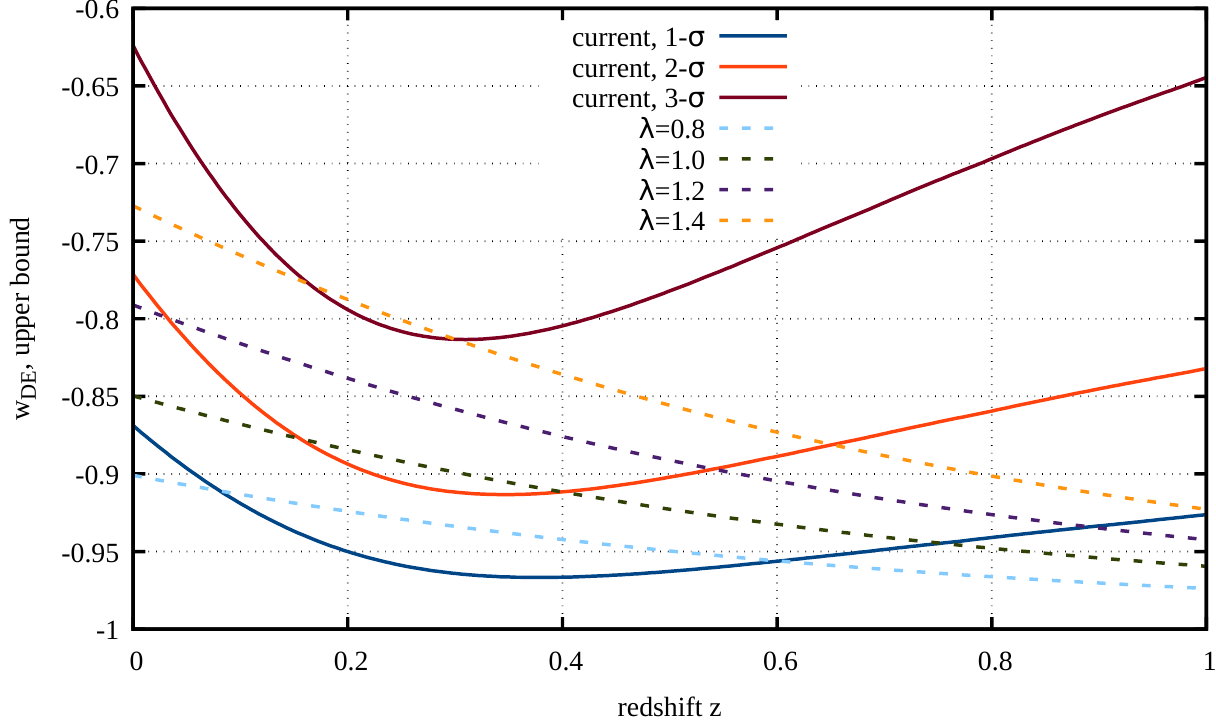}
\caption{The figure shows 1- and 2-$\sigma$ upper bounds on the reconstructed equation of state $w(z)$ as a function of redshift, according to the analysis described in the text. We use the constraints obtained from SNeIa, CMB, BAO and $H_0$ measurements shown in Fig.\ 21 of \cite{Scolnic:2017caz} (the yellow contours there) combined with the CPL parameterization of $w(z)$. We find that our 1-$\sigma$ bound reproduces the tight constraint addressed as a 3-$\sigma$ upper bound in \cite{Agrawal:2018own}. We also extrapolate the uncertainty in order to plot the 3-$\sigma$ upper bound. The dotted lines show the equation of state as a function of redshift for different quintessence models with different $\lambda$ values for comparison.}
\label{fig_wDEvsz}
\end{figure}

We proceed in the following way. Let $M$ be the covariance matrix (or, the inverse Fisher matrix) obtained from observational constraints on $w_0$ and $w_a$, and let $\vec w = (w_0, w_a)^\top$ be a vector composed of these two parameters. Then 
\begin{equation}
  \vec w^\top M\vec w = 1
\label{ellipse}
\end{equation} 
defines an ellipse enclosing the domain in the $(w_0, w_a)$ plane allowed by the observation. Inserting $\vec w(\varphi) = r_w(\cos\varphi, \sin\varphi)^\top$ for $\varphi\in[0,2\pi]$ into (\ref{ellipse}) defines the radius $r_w(\varphi)$ tracing this contour, from which
\begin{equation}
  w_0(\varphi) = r_w(\varphi)\cos\varphi\;,\quad
  w_a(\varphi) = r_w(\varphi)\sin\varphi
\end{equation}
can be read off. Searching for the maximum of $w(z)$ according to the CPL parameterization along this contour,
\begin{equation}
  w_\mathrm{max}(z) :=
  \max_\varphi\left[w_0(\varphi)+\frac{z}{1+z}w_a(\varphi)\right]\;,
\end{equation}
returns the upper bound on the values of $w(z)$ compatible with the observational constraints. Figure \ref{fig_wDEvsz} shows the upper bounds obtained from the 1- and 2-$\sigma$ contours given by \cite{Scolnic:2017caz}.

In this way, we can directly map an elliptical approximation to the contours outlining observational uncertainties via an analytical expression to an upper bound on the equation-of-state parameter, using the CPL parametrisation. The observational uncertainties on $w_0$ and $w_a$ can then be compared with the regime still allowed by the string Swampland criteria.

This is illustrated in Fig.\ \ref{fig_wDEvsz}. We see there that our 2-$\sigma$ contour reaches only a minimum of $w\approx-0.91$, whereas \cite{Agrawal:2018own} find a minimum of $w\approx-0.96$ even at the 3-$\sigma$ level. 
We will apply the method outlined above to construct analytically an upper bound on the equation of state $w(z)$ from current and future observational constraints. We discard equation-of-state parameter values smaller than $-1$, since cosmological perturbations are ill-defined in this case.

Fig.\ \ref{fig_wDEvsz} also shows the equation-of-state $w(z)$ for different values of $\lambda$, obtained by solving Eqs.\ \eqref{EOMquintessence_a} and \eqref{EOMquintessence_b} numerically for $c=\lambda$. The current 2-$\sigma$ cosmological constraints allow quintessence models with $\lambda \lesssim 0.9$. Even though the 3-$\sigma$ contours are not shown in Fig.\ 21 of \cite{Scolnic:2017caz}, we can extrapolate the uncertainty to only restrict $\lambda$ to be $\lambda \lesssim 1.35$ (also shown in Fig.\ \ref{fig_wDEvsz}). This is in agreement with the string Swampland criteria $|V'|/V>c=\lambda\sim\mathcal{O}(1)$. We will later see in Fig.\ \ref{fig_future} that the dark-energy models with $\lambda\le0.1$ can barely be distinguished from the $\Lambda$CDM model with $w=-1$. This would be in great tension with the string Swampland criterion and thus push string theory into an uncomfortable corner. However, the current 3-$\sigma$ constraints with $c=\lambda \lesssim 1.35$ are still very far from this constraining power.

\section{Future prospects}

As we can see in Fig.\ \ref{fig_wDEvsz}, the current 2-$\sigma$ observational constraints on the quintessence models allow models with $\lambda \lesssim 0.9$, which is still well in agreement with the string Swampland criteria. We will now investigate whether these constraints can be significantly improved by future surveys, and whether future observations may ever be able to distinguish between a pure cosmological constant and dark-energy model, and hence drive string theory into a corner.

It is well known that the search for new physics at LHC requires an increase of the collision energy and of the accelerator luminosity. Analogously, cosmic surveys pursue new physics by increasing their redshift limits (their survey volume) and the number of observed objects. This is the main goal of Stage-4 surveys such as DESI, LSST, Euclid, and others \cite{stage4}. One important outcome of these surveys will be strong constraints on the equation of state of dark energy. In return, such constraints will tightly restrict the allowed range of dark-energy models.

\begin{figure}[h!]
  \includegraphics[width=\hsize]{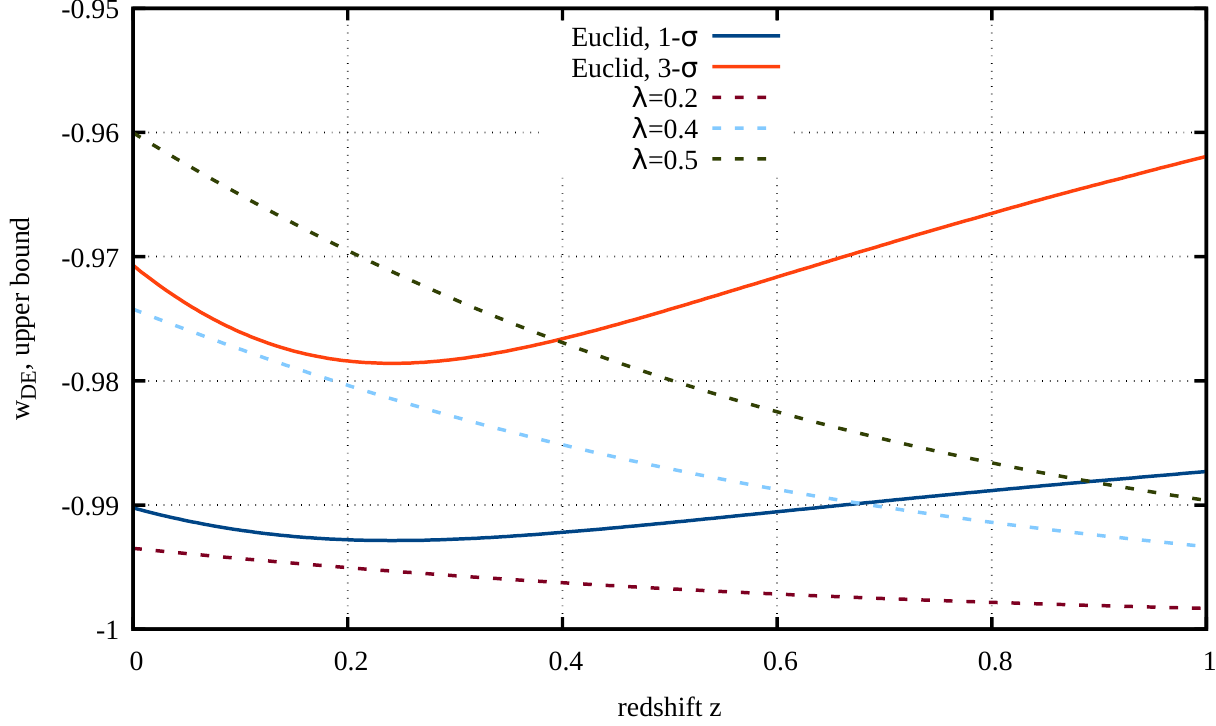}
\caption{Similar to Fig.\ \ref{fig_wDEvsz}, upper bounds on $w(z)$ are shown here as expected from the target uncertainties in $w_0$ and $w_a$ given in Tab.\ 2.2 of the Euclid Red Book. At the 3-$\sigma$ level, $\lambda$ will be constrained to $\lambda\lesssim0.4$.}
\label{fig_euclid}
\end{figure}

In order to illustrate the near-future limits that Stage-4 surveys might obtain (Euclid's launch is planned for 2020), we show in Fig.\ \ref{fig_euclid} the prospective 1- and 3-$\sigma$ upper bounds on $w(z)$. We have obtained them by adopting the orientation of the inverse Fisher matrix between $w_0$ and $w_a$ illustrated in \cite{Scolnic:2017caz} and the limits on $w_0$ and $w_a$ given in Tab.\ 2.2 of the Euclid Definition Study Report (the ``Red Book'', \cite{euclidRedBook}). Since the inverse Fisher matrix for $(w_0, w_a)$ to be expected from Euclid is not publicly available yet, we assumed its orientation in the $(w_0, w_a)$ plane to follow that shown by \cite{Scolnic:2017caz}. Changes in the orientation angle of the Fisher ellipse do not strongly affect the constraints derived.

As we can see in Fig.\ \ref{fig_euclid}, the projected limits of Stage-4 surveys will change the situation substantially. Following the 3-$\sigma$ limits from the Euclid Red Book, we find the forecasted upper limit $w(z) \lesssim -0.97$ for the equation of state in the redshift interval $0\le z\le 0.6$. The tighter constraints of Euclid on the other hand drive the allowed dark-energy models to $\lambda\lesssim 0.4$. Compared to the current 3-$\sigma$ constraints of $\lambda\lesssim 1.35$, this will be a significant reduction.

This improvement by a factor of 3 on the one hand will remove a large class of quintessence models. On the other hand, the surviving quintessence models would be pushed into notable tension with the string Swampland criteria and could therefore also be discarded for dark energy phenomenology. Euclid aims at measuring the redshift of galaxies up to $z\approx2$. For redshifts $z>1$, the CPL parametrisation may not be adequate any more, and a more adaptable model may further improve the constraints on the Swampland criteria.

\begin{figure}[h!]
  \includegraphics[width=\hsize]{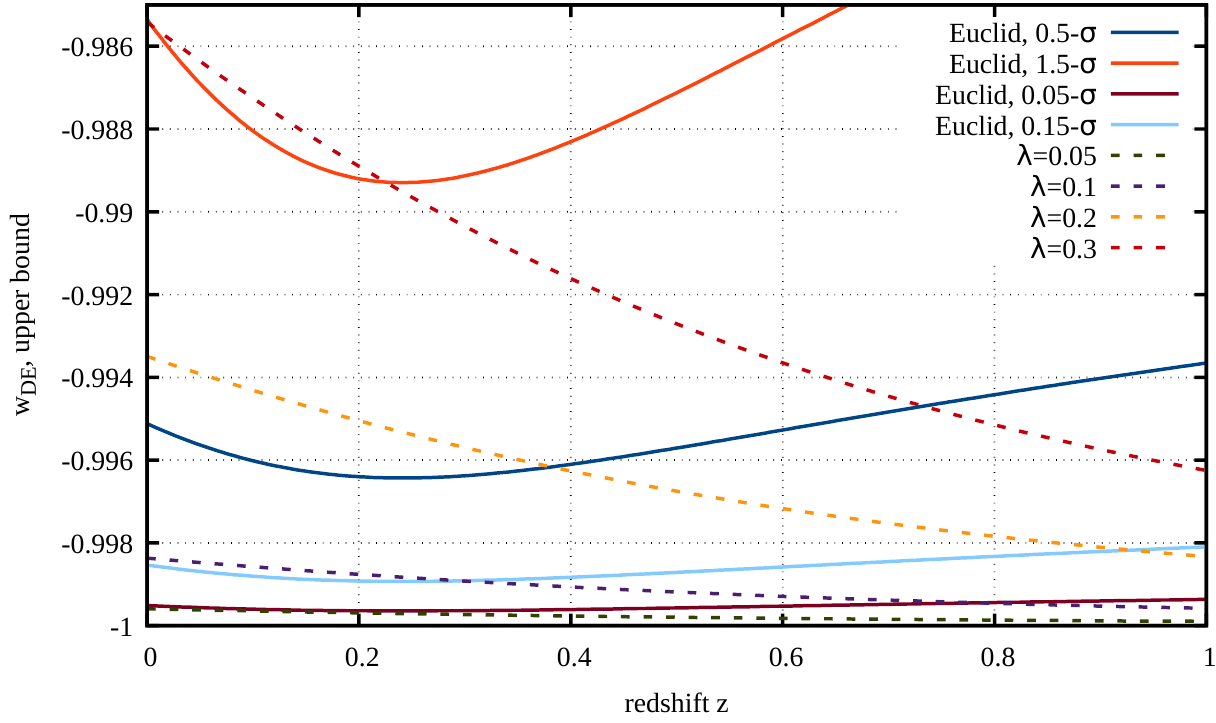}
\caption{Upper bounds on $w(z)$ to be expected from a hypothetical future survey with uncertainties on $w_0$ and $w_a$ lowered by 50\% compared to the values given in the Euclid Red Book. At the 3-$\sigma$ level, only $\lambda\lesssim0.3$ will be allowed, which would get into tension with the Swampland criteria. Constraining $\lambda$ down to $\lambda\lesssim0.1$ is only achievable if the uncertainties could be reduced by a factor of $\sim20$ compared to Euclid.}
\label{fig_future}
\end{figure}

As an outlook on possible future constraints beyond Stage-4 surveys, we show similar curves in Fig.\ \ref{fig_future}. We obtained them by assuming that the uncertainties on $w_0$ and $w_a$ could further be reduced by 50\% compared to values given in the Euclid Red Book. A hypothetical future survey with uncertainties in $w_0$ and $w_a$ lowered by half would significantly reduce $w(z)$ down to $-0.985$ between redshifts $0\le z\le 0.6$, leaving room for only very small deviations from a cosmological constant. Surviving quintessence models would be restricted to have $\lambda\lesssim0.3$, which would get into notable tension with the Swampland criteria. 

How severely could idealized futuristic surveys constrain dark-energy models? An extrapolation of Euclid's uncertainties reveals that constraining $\lambda$ to $\lambda\lesssim0.1$ could only be achieved if the uncertainties of $w_0$ and $w_a$ could be reduced by a factor of $\sim20$ compared to Euclid or, in other words, if the survey volume could be further enlarged by a factor of $\sim400$ beyond Euclid's. In view of such perspectives, it may appear doubtful that such an accuracy may ever be reached. However, already in the near future, Stage-4 surveys and their followers will push limits on $\lambda$ down to $\lambda\lesssim 0.4$, which would be in uncomfortable tension with the Swampland criteria.

\section{Conclusion}

In this Letter, we have investigated the implications of the string Swampland criteria for dark-energy models based on a scalar field in view of ongoing and future cosmological observations. Should string theory be the ultimate quantum gravity theory, then, according to the second of the swampland conjectures, de Sitter solutions with a positive cosmological constant cannot account for the fate of the late-time universe. 
Dark-energy models based on scalar fields evolving in time still have to satisfy certain criteria in order to remain outside the Swampland \footnote{Note that string theory does not naturally lead to scalar fields with the energy scale required to be a candidate for quintessence. Rather, string theory may require radically new ideas to explain dark energy.} . We have studied the observational implications of future surveys on such quintessence models and the associated restrictions on the fate of dark energy.

Current limits on $w_0$ and $w_a$ impose already a quite tight upper bound on $w(z)$. According to our analysis, current 2-$\sigma$ limits demand that $w(z) \lesssim -0.91$ at $z \approx 0.3$. This constrains the parameter $\lambda$ to $\lambda \lesssim 0.9$, which is still well in agreement with the string Swampland criteria. The projected limits to be obtained from Stage-4 surveys can be expected to change the situation substantially. Applying 3-$\sigma$ limits taken from the Euclid Red Book, we find that an upper bound on $w(z)$ will be $w(z) \lesssim -0.97$ in the redshift interval $0\le z\le 0.6$. This will restrict the allowed range for $\lambda$ to $\lambda\lesssim 0.4$. Surviving quintessence models would thus be driven into substantial tension with the string Swampland criteria. Should a future survey be able to lower the uncertainties in $w_0$ and $w_a$ to half the uncertainties expected from Euclid, $w(z)$ would have to fall below $-0.985$ in the redshift range $0\le z\le 0.6$ and thus only leave room for small deviations of quintessence from a cosmological constant. The parameter $\lambda$ would then have to fall below $\lambda\lesssim0.3$.

These possibly idealised constraints raise the question of how tightly any future survey will ever be able to 
 differentiate between whether the dark energy is a cosmological constant or not. We estimated that constraining $\lambda$ to $\lambda\lesssim0.1$ would require a reduction in the uncertainties of $w_0$ and $w_a$ compared to those expected from Euclid by a factor of $\sim20$. This would correspond to increasing the survey volume by a factor of $\sim400$ compared to that covered by Euclid. In view of such perspectives, there will be fundamental observational limitations on lowering $\lambda$ to $\lambda\lesssim0.1$ with future surveys. 

%
\section*{Acknowledgements}
We are grateful for useful discussions with Prateek Agrawal, Elisa Ferreira, Elena Sellentin, Paul J. Steinhardt, and Cumrun Vafa, and comments by Arthur Hebecker and Savdeep Sethi. LH thanks financial support from Dr.~Max R\"ossler, the Walter Haefner Foundation and the ETH Zurich Foundation.  RB is supported in part by funds from NSERC and from the Canada Research Chair program.


\end{document}